\documentclass[aps,prd,onecolumn,preprintnumbers,nofootinbib,superscriptaddress]{revtex4-2}
\input{header.tex}

\usepackage{fontawesome5}
\definecolor{color_git}{rgb}{0.098, 0.160, 0.345}
\newcommand{\gitlink}{\href{https://github.com/KensukeAkita/sterile-dm-lfa/tree/main}{\textsc{g}it\textsc{h}ub {\large\color{color_git}\faGithub}}}

\definecolor{forestgreen}{RGB}{34,139,34}

\newcommand{\MP}{{M_{\rm P}}}

\begin{document}

\reportnum{-2}{CERN-TH-2025-185}

\title{
Affleck-Dine Leptoflavorgenesis
}

\author{Kensuke~Akita}
\email{kensuke@hep-th.phys.s.u-tokyo.ac.jp}
\affiliation{Department of Physics, The University of Tokyo, Bunkyo-ku, Tokyo 113-0033, Japan}
\author{Koichi Hamaguchi}
\email{hama@hep-th.phys.s.u-tokyo.ac.jp}
\affiliation{Department of Physics, The University of Tokyo, Bunkyo-ku, Tokyo 113-0033, Japan}
\author{Maksym~Ovchynnikov}
\email{maksym.ovchynnikov@cern.ch}
\affiliation{Theoretical Physics Department, CERN, 1211 Geneva 23, Switzerland}

\date{\today}

\begin{abstract}
We propose a scenario to produce large primordial lepton flavor asymmetries with vanishing
total lepton asymmetry, based on the Affleck-Dine mechanism with Q-ball formation.
This scenario can produce large lepton flavor asymmetries
while automatically maintaining the vanishing total lepton number without fine-tuning, evading the current BBN and the CMB constraints by neutrino oscillations at MeV temperature. The asymmetries can be produced at cosmic temperatures of $T\gtrsim 1\ {\rm GeV}$, early enough to have broad impacts from the early Universe to the present cosmology. 
This scenario could affect various aspects of early Universe cosmology simultaneously or separately: (i) explaining the observed baryon asymmetry by the same origin as the lepton flavor asymmetries, (ii) affecting the nature of the QCD transition,
(iii) opening up a new parameter space of sterile neutrino dark matter by enhancing their production, and (iv) altering the abundance of the light elements, in particular, resolving the recently reported helium-4 anomaly. 
\end{abstract}

\maketitle

\tableofcontents

\noindent\makebox[\linewidth]{\rule{0.8 \paperwidth}{0.2pt}}
\vspace{1 mm}
\vspace{-4.5mm}

\section{Introduction}
\label{sec:introduction}

We observe that the Universe consists almost entirely of matter and a little antimatter. 
This is commonly parameterized as the number density of baryons minus antibaryons normalized to the entropy density,
given by $Y_B\equiv \Delta n_B/s= (n_B-n_{\bar{B}})/s=(8.75 \pm 0.23)\times 10^{-11}$ \cite{Planck:2018vyg}. 

On the other hand, we have less clues about primordial lepton asymmetries, $Y_{L_\alpha}\equiv\Delta n_{L_\alpha}/s = (n_{L_\alpha}-n_{\bar{L}_\alpha})/s$, where $n_{L_\alpha}$ ($n_{\bar{L}_\alpha}$) is the number densities of the (anti)lepton with flavor $\alpha$ $(\alpha = e,\ \mu,\ \tau)$.
Naively, baryon and lepton asymmetries would have similar magnitudes, $Y_B\sim Y_{L_\alpha}$, due to the electroweak sphaleron transition in the early Universe \cite{Kuzmin:1985mm}. However, the sphaleron processes are ineffective at the temperature of the Universe of $T\lesssim T_{\rm sph}\simeq 130\ {\rm GeV}$ \cite{DOnofrio:2014rug}, and large lepton asymmetry can be produced below this temperature, consistent with the observed small baryon asymmetry. Indeed, scenarios to produce large primordial lepton asymmetry have been proposed in refs.~\cite{Dolgov:1989us,Casas:1997gx,March-Russell:1999hpw,Kawasaki:2002hq,Yamaguchi:2002vw,Pilaftsis:2003gt,Shaposhnikov:2008pf,Shu:2006mm,Gu:2010dg,Kawasaki:2022hvx,Borah:2022uos,Kasai:2024diy,Borah:2024xoa}. 
Observationally, the Big Bang Nucleosynthesis (BBN) and the Cosmic Microwave Background (CMB) are sensitive to lepton asymmetries \cite{Sarkar:1995dd,Iocco:2008va,Pitrou:2018cgg,Lesgourgues:2006nd,Serpico:2005bc,Mangano:2011ip,Chu:2006ua,Simha:2008mt,Li:2024gzf,Oldengott:2017tzj,Seto:2021tad,Kumar:2022vee,Matsumoto:2022tlr,Burns:2022hkq,Escudero:2022okz,Froustey:2024mgf,Domcke:2025lzg,Yanagisawa:2025mgx}. The former primarily probes the asymmetry in the electron neutrino flavor through the neutron-proton ratio. In addition, both the BBN and the CMB are sensitive to the total radiation density enhanced by the asymmetries.

In particular, there are almost no clues about lepton flavor asymmetries summing up to zero total lepton asymmetry, as recently studied comprehensively in Ref.~\cite{Domcke:2025lzg}. The BBN and CMB observations are sensitive to the asymmetries
below the temperature of the Universe of $T\simeq 1\ {\rm MeV}$.
At $T\sim 15\ {\rm MeV}$ before neutrino decoupling, neutrino oscillations start \cite{Dolgov:2002ab,Pastor:2008ti,Mangano:2010ei,Castorina:2012md,Froustey:2021azz,Froustey:2024mgf,Domcke:2025lzg}. The flavor-changing oscillations can lead to washing out the flavor asymmetries, relaxing the BBN and CMB constraints on them.

Large lepton asymmetry can induce rich phenomenology in the Universe. In particular, less constrained lepton flavor asymmetries might affect baryogenesis \cite{Casas:1997gx,Dolgov:1989us,March-Russell:1999hpw,Kawasaki:2002hq,Yamaguchi:2002vw,Pilaftsis:2003gt,Shaposhnikov:2008pf,Shu:2006mm,Gu:2010dg,Kawasaki:2022hvx,Borah:2022uos,Borah:2024xoa,Cohen:1987vi,Cohen:1988kt,Davidson:1994gn,Chiba:2003vp,Takahashi:2003db,Gao:2023djs,Gao:2024fhm}, the nature of the QCD transition \cite{Asakawa:1989bq,Schwarz:2009ii,Middeldorf-Wygas:2020glx,Vovchenko:2020crk,Gao:2021nwz,Ferreira:2025zeu,Formaggio:2025nde}, the production of sterile neutrino dark matter \cite{Shi:1998km,Abazajian:2001nj,Kishimoto:2008ic,Laine:2008pg,Ghiglieri:2015jua,Venumadhav:2015pla,Kasai:2024diy,Gorbunov:2025nqs,Vogel:2025aut,Akita:2025txo}, the detection prospects of the cosmic neutrino background \cite{Stodolsky:1974aq,Duda:2001hd,Long:2014zva,Ruchayskiy:2022eog}, the presence of cosmic magnetic fields \cite{Joyce:1997uy,Boyarsky:2011uy,Akamatsu:2013pjd,Hirono:2015rla,Rogachevskii:2017uyc,Domcke:2022uue} and the abundance of the light elements \cite{Kohri:1996ke,Casas:1997gx,March-Russell:1999hpw,Kawasaki:2002hq,Matsumoto:2022tlr,Kawasaki:2022hvx,Escudero:2022okz,Froustey:2024mgf,Domcke:2025lzg,Burns:2022hkq}.
Primordial lepton asymmetries may also resolve the recently reported helium-4 anomaly, indicating a smaller helium-4 abundance compared to the standard BBN scenario~\cite{Matsumoto:2022tlr,Escudero:2022okz,Froustey:2024mgf,Domcke:2025lzg,Burns:2022hkq,Yanagisawa:2025mgx}.
This anomaly favors a large electron-flavor asymmetry at the BBN epoch in a mild $2\sigma$ tension with the standard scenario \cite{Yanagisawa:2025mgx}.

The sphaleron process preserves the quantity $Y_B/3-Y_{L_\alpha}$ for each lepton flavor $\alpha$ but can change 
$Y_B + Y_L$, where $Y_L=\sum_{\alpha}Y_{L_\alpha}$ is the total lepton asymmetry. This implies that the baryon asymmetry vanishes if $Y_B-Y_L=0$. Interestingly, however, the sphaleron conversion from lepton flavor asymmetries with $Y_B-Y_L=0$ to baryon asymmetry is not perfectly canceled due to the differences in the Standard Model (SM) lepton Yukawa couplings~\cite{Khlebnikov:1988sr,March-Russell:1999hpw,Laine:1999wv}.
Large lepton flavor asymmetries then may offer a natural explanation of small baryon asymmetries while providing a rich phenomenology for the history of the Universe.

One of the open questions about large lepton flavor asymmetries is their origin: What is the leptoflavorgenesis scenario? Until now, several models for leptoflavorgenesis have been proposed \cite{Kuzmin:1987wn,March-Russell:1999hpw,Shu:2006mm,Gu:2010dg,Mukaida:2021sgv}. However, it would be non-trivial how large asymmetries leptoflavorgenesis scenarios can produce under the latest BBN and CMB constraints. In addition, it would also be non-trivial whether the asymmetries are generated early enough to affect the early Universe physics.

In this work, we propose a new scenario to produce large lepton flavor asymmetries, based on the Affleck-Dine (AD) mechanism \cite{Affleck:1984fy,Dine:1995kz}.
We utilize the $Q\bar{u}L\bar{e}$ flat directions, which can break lepton flavor but conserve the total lepton number.
To prevent the overproduction of the baryon asymmetry, we consider a scenario of Q-ball formation~\cite{Kawasaki:2002hq,Kawasaki:2022hvx,Kasai:2024diy}, where the generated lepton flavor asymmetries are confined within the Q-balls and protected from being converted into baryon asymmetry. 
The lepton flavor asymmetries are then produced via the late-time decays of the Q-balls.
This Affleck-Dine leptoflavorgenesis (ADLFG) scenario can produce large lepton flavor asymmetries 
while automatically maintaining the vanishing total lepton number without fine-tuning, evading the BBN and CMB constraints. The asymmetries can be produced at the temperature of the Universe of $T\gtrsim 1\ {\rm GeV}$, which is early enough to affect the nature of the QCD transition and the production of sterile neutrino dark matter. 

Our main results are summarized in Figure~\ref{fig:parameterspace_L}.
This ADLFG scenario can generate large lepton flavor asymmetries of $|Y_{L_\alpha}|\sim 10^{-3}\cdots10^{-1}$. If tau flavor asymmetry of $Y_{L_\tau}\sim-10^{-2}\cdots-10^{-1}$ or muon flavor asymmetry of $Y_{L_\mu}\sim -10^{-1}$ is produced, the observed baryon asymmetry can be generated by the same origin as the lepton flavor asymmetries in this scenario (cf.~Eq.~\eqref{eq:YB_vs_YLa}).
The generated lepton flavor asymmetries can open up a new parameter space of sterile neutrino DM \cite{Akita:2025txo}, alter the abundance of the light elements, in particular, resolve the recently reported helium-4 anomaly \cite{Domcke:2025lzg}, and may affect the nature of the QCD transition~\cite{Asakawa:1989bq,Schwarz:2009ii,Middeldorf-Wygas:2020glx,Vovchenko:2020crk,Gao:2021nwz,Ferreira:2025zeu,Formaggio:2025nde} as discussed in Section~\ref{sec:implication}.

The rest of this paper is organized as follows.
In Sec.~\ref{sec:flat directions}, we introduce lepton flavor-violating flat directions in the minimal supersymmetric standard model (MSSM), which is utilized in the AD mechanism.
In Sec.~\ref{sec:ADLFG_with_Q}, we propose an ADLFG scenario to generate large lepton flavor asymmetries. 
Sec.~\ref{sec:implication} presents implications of the proposed scenarios for the early Universe cosmology.
Sec.~\ref{sec:conclusions} concludes the paper.
Appendix~\ref{app:list_of_paras} provides a list of input parameters and key derived quantities used in our analysis. Appendix~\ref{app:Ndof}, \ref{app:evap-diff}, and \ref{app:Gravitino_from_Q} discuss Q-ball decay channels, Q-ball evaporation, and Q-ball decay into gravitinos in our scenario, respectively.


\section{Lepton flavor violating flat directions}
\label{sec:flat directions}

The AD mechanism utilizes flat directions in the minimal supersymmetric standard model (MSSM), along which there are no classical potentials at the renormalizable level. Squarks and/or sleptons condense along flat directions. If a flat direction violates lepton flavor symmetry but conserves $B-L$ charge, large lepton flavor asymmetries may be produced. In this section, we show that such lepton-flavor-violating flat directions exist in the MSSM.  

The MSSM superpotential is given by
\begin{align}
    W_{\rm MSSM}=
    y^u_\alpha Q_\alpha\bar{u}_\alpha H_u
    -y^d_{\alpha\beta}Q_\alpha\bar{d}_\beta H_d
    -y^e_\alpha L_\alpha \bar{e}_\alpha H_d 
+\mu H_uH_d,
\label{eq:WMSSM}
\end{align}
where $Q_\alpha$, $\bar{u}_\alpha$, $\bar{d}_\beta$, $L_\alpha$,  $\bar{e}_\alpha$, $H_u$ and $H_d$, denote the superfields of the left-handed quarks, right-handed up quarks, right-handed down quarks, left-handed leptons, right-handed leptons, up-type Higgs, and down-type Higgs, respectively, with $\alpha,\beta=1,2,3$ being the flavor indices.
Here and in the following, without loss of generality, we work in a basis where the Yukawa matrices for the up-type quarks and charged leptons are diagonal.

The flat directions in the MSSM are parametrized by gauge-invariant polynomials of chiral superfields~\cite{Gherghetta:1995dv}. Among them, we adopt the $Q\bar{u}L\bar{e}$ directions, which 
correspond to the lowest-dimensional gauge-invariant polynomials that (i) can break lepton flavor, (ii) conserve the total lepton number, and (iii) remain unlifted by the renormalizable superpotential.\footnote{In Ref.~\cite{March-Russell:1999hpw}, the authors considered the flat direction of $|L_e|^2+|L_\mu|^2=|H_u|^2$ with a K\"ahler potential $K\propto L_e^\dagger L_\mu H_u^\dagger H_u$. While this is also a viable possibility, it requires the superpotential for the neutrino masses, $W\sim (L_\alpha H_u)(L_\beta H_u)$, to vanish in order to conserve the total lepton number during the AD mechanism, implying that an alternative origin for neutrino masses must be considered. For this reason, we do not adopt this scenario in the present work. Note that such lepton-number-violating superpotentials do not lift the $Q\bar{u}L\bar{e}$ flat direction and hence they are compatible with our scenario.
}
Such flat directions are expressed in terms of the AD field $\phi$ as
\begin{align}
Q_\alpha&=\frac{1}{2}\begin{pmatrix}
\phi & 0 & 0  \\  0 & 0 & 0
\end{pmatrix},\quad
\bar{u}_\beta=\frac{1}{2}\begin{pmatrix}
\phi & 0 & 0 
\end{pmatrix},
\quad (\alpha\ne \beta),
\notag\\
L_\gamma&=\frac{1}{2}\begin{pmatrix}
0  \\ \phi
\end{pmatrix},\quad
\bar{e}_\delta=\frac{1}{2}\phi,\quad (\gamma\ne \delta),
\end{align}
where the rows and columns correspond to the SU(2)$_L$ and SU(3)$_C$ components, respectively, and the superfields and their scalar components are represented by the same symbols, for simplicity.
The $F$-flatness conditions require that the flavor indices satisfy $\alpha\ne \beta$ and $\gamma\ne \delta$, which ensures that the MSSM superpotential in Eq.~\eqref{eq:WMSSM} does not generate a quartic potential for $\phi$.
For simplicity, we fix the quark sector flavors to, e.g., $(\alpha,\beta)=(2,3)$, and omit those indices in what follows.
We assume that the third-generation squark is included so that the Q-balls are not formed by the gravity-mediation potential (see discussion below).

The lepton flavor associated with the flat direction is important, since it determines the generated lepton flavor asymmetry. Omitting the quark indices, the most general $Q\bar{u}L\bar{e}$ directions are parametrized as 
\begin{align}
Q\bar{u}L_\alpha\bar{e}_{\beta\gamma} &\text{ direction }
(\alpha\ne \beta,\; \alpha\ne \gamma)
\notag\\
Q&=\frac{1}{2}\begin{pmatrix}
\phi & 0 & 0  \\  0 & 0 & 0
\end{pmatrix},\;
\bar{u}=\frac{1}{2}\begin{pmatrix}
\phi & 0 & 0 
\end{pmatrix},\;
L_\alpha=\frac{1}{2}\begin{pmatrix}
0  \\ \phi
\end{pmatrix},\;
\bar{e}_\beta=\frac{\cos\chi}{2}\phi,\;
\bar{e}_\gamma=\frac{\sin\chi}{2}\phi\,,
\label{eq:QuLe}\\
Q\bar{u}L_{\alpha\beta}\bar{e}_\gamma &\text{ direction }
(\alpha\ne \gamma,\; \beta\ne \gamma)
\notag\\
Q&=\frac{1}{2}\begin{pmatrix}
\phi & 0 & 0  \\  0 & 0 & 0
\end{pmatrix},\;
\bar{u}=\frac{1}{2}\begin{pmatrix}
\phi & 0 & 0 
\end{pmatrix},\;
L_\alpha=\frac{\cos\chi}{2}\begin{pmatrix}
0  \\ \phi
\end{pmatrix},\;
L_\beta=\frac{\sin\chi}{2}\begin{pmatrix}
0  \\ \phi
\end{pmatrix},\;
\bar{e}_\gamma=\frac{1}{2}\phi,
\end{align}
which are $F$- and $D$-flat at the renormalizable level.
In the following, to avoid unnecessary case-by-case distinctions, we concentrate on the case of the $Q\bar{u}L_\alpha \bar{e}_{\beta\gamma}$ directions. The results for the $Q\bar{u}L_{\alpha\beta} \bar{e}_\gamma$ directions can be obtained in a straightforward manner.
Along these flat directions, the total lepton number always remains zero. 
The $F$-flatness ensures that the flavors of the $L_\alpha$ and $\bar{e}_\beta$, $\bar{e}_\gamma$ are different, i.e., $\alpha\ne \beta,\gamma$. On the other hand, the angle $\chi$, which determines the relative contribution of the $\beta$- and $\gamma$-flavored components, is not fixed by the flatness conditions. In general, $\chi$ evolves during the AD mechanism and its dynamics can be complicated (see, e.g., Ref~\cite{Enqvist:2003pb}). In this work, for simplicity, we do not follow the time evolution of $\chi$, and instead parametrize the final flavor asymmetry by a fixed value of $\chi$.

As we discuss in the next section, the AD mechanism generates a non-zero $\phi$-number:
\begin{align}
n_\phi = i\left(\dot{\phi}^* \phi - \phi^* \dot{\phi}\right)=2\dot{\theta}|\phi|^2,
\label{eq:phi_number}
\end{align}
where the dot denotes the derivative with respect to the cosmic time $t$ and $\theta$ is the phase of the AD field $\phi$.
This results in lepton flavor asymmetries:
\begin{align}
\Delta n_{L_\alpha}=\frac{\kappa_\alpha}{4}n_\phi,
\label{eq:nLi_nphi}
\end{align}
where
\begin{align}
(\kappa_\alpha, \kappa_\beta, \kappa_\gamma) = 
(1,-\cos^2\chi,-\sin^2\chi),
\label{eq:kappa_and_chi}
\end{align}
for the $Q\bar{u}L_\alpha \bar{e}_{\beta\gamma}$ direction.
For example, for $\chi=0$, the AD mechanism along the $Q\bar{u}L_e \bar{e}_\mu$ direction can generate $L_e-L_\mu$ asymmetry, $\Delta n_{L_e}=-\Delta n_{L_\mu}$.
Another direction, $Q\bar{u}L_e\bar{e}_{\mu\tau}$ direction with $\chi \simeq \pi/4$, leads to $(\Delta n_{L_e}, \Delta n_{L_\mu}, \Delta n_{L_\tau})\simeq (1,-1/2,-1/2)(n_\phi/4)$, which will be discussed in Sec.~\ref{sec:implication} as a benchmark point.


\section{Affleck-Dine leptoflavorgenesis 
}
\label{sec:ADLFG_with_Q}

We examine the generation of large lepton flavor asymmetries via the AD mechanism along the $Q\bar{u}L\bar{e}$ direction introduced in the previous section.
The generated lepton flavor asymmetries and the baryon asymmetry of the Universe (BAU) depend on whether the AD fields experience spatial instabilities during their coherent oscillations, forming non-topological solitons called Q-balls \cite{Coleman:1985ki,Kusenko:1997si,Enqvist:1997si,Kasuya:1999wu}. 
To prevent the overproduction of the baryon asymmetry, we consider a scenario of Q-ball formation, where the generated lepton flavor asymmetries are confined within the Q-balls and protected from being converted into baryon asymmetry. 
The main assumptions and setup of our scenario are as follows:
\begin{itemize}
\item We utilize the $Q\bar{u}L\bar{e}$ flat directions, which break lepton flavor while conserving the total lepton number.
\item We assume the absence of nonrenormalizable superpotentials lifting the AD flat direction, which allows for a large initial amplitude of the AD field.
Lepton flavor asymmetries are generated through higher-dimensional operators in the K\"ahler potential.
\item 
We assume a gauge-mediated SUSY breaking scenario where delayed-type Q-balls~\cite{Kasuya:2001hg} form. 
These Q-balls confine the generated lepton flavor asymmetries and protect them from being converted into baryon asymmetry~\cite{Kawasaki:2002hq,Kawasaki:2022hvx,Kasai:2024diy}.
The lepton flavor asymmetries are then produced via the late-time decays of the Q-balls.
\end{itemize}

\subsection{Large lepton flavor asymmetries from delayed-type Q-balls}

We consider the gauge-mediated SUSY breaking scenario with Q-ball formation. In the gravity-mediation case with Q-ball formation, the late-time decay of Q-balls typically leads to an overproduction of the lightest supersymmetric particles (LSPs), unless their annihilation cross section is sufficiently large~\cite{Fujii:2002kr} (cf.~\cite{Kamada:2012bk}).
A detailed analysis of that case is left for future work.

The scalar potential of the AD field $\phi$ in Eq.~\eqref{eq:QuLe} is induced by nonrenormalizable terms in super- and K\"ahler potential. In this work, 
we are interested in the generation of large lepton flavor asymmetries, which, as shown in Eq.~\eqref{eq:phi_number}, requires a large $\phi$ field amplitude. Such a large amplitude can be realized if the AD field is not lifted by nonrenormalizable superpotentials. We therefore assume that the superpotential does not contain such terms, and instead consider the following term in the K\"ahler potential:
\begin{align}
K = \lambda_{\alpha\beta} \frac{Z Z^\dagger}{\MP^2}  Q\bar{u}L_\alpha\bar{e}_\beta + {\rm h.c.}\,,
\label{eq:Kahler}
\end{align}
where $\lambda_{\alpha\beta}$ is a dimensionless coupling, $\MP=2.4\times 10^{18}~{\rm GeV}$ is the reduced Planck scale, and $Z$ is the superfield whose non-vanishing $F$-term is responsible for the SUSY breaking.

The scalar potential for the AD field $\phi$ is given by
\begin{align} 
    V_0(\phi)&=V_{\rm gauge}(\phi)+V_{\rm grav}(\phi)+\left(\frac{\lambda m_{3/2}^2}{4\MP^2}\phi^4+{\rm h.c.}\right)
+\cdots\,, \label{eq:V0}
\end{align}
where the third term in eq.~\eqref{eq:V0}, which originates from the K\"ahler potential Eq.~\eqref{eq:Kahler}, violates lepton flavor asymmetries, and $ m_{3/2}$ denotes the gravitino mass.
The ellipsis indicates higher-order terms suppressed by the Planck scale, which are expected to exponentially lift the scalar potential above the Planck scale due to supergravity effects.
We take $\lambda\sim \lambda_{\alpha\beta}$
to be real and positive through field redefinitions.
$V_{\rm gauge}$ denotes the potential induced by the gauge-mediated SUSY breaking effect \cite{deGouvea:1997afu,Kusenko:1997si,Kasuya:2001hg},
\begin{align}
    V_{\rm gauge}(\phi)=
    \begin{cases}
      m_\phi^2|\phi|^2  & (|\phi| \ll M_S)\,,  \\
       M_F^4\left(\log\frac{|\phi|^2}{M_S^2} \right)^2 & (|\phi| \gg M_S)\,,
    \end{cases}
    \label{eq:Vgauge}
\end{align}
where $m_\phi^2=(
m_{\tilde{Q}}^2
+m_{\tilde{\bar{u}}}^2
+m_{\tilde{L}_\alpha}^2
+m_{\tilde{\bar{e}}_\beta}^2 \cos^2\chi
+m_{\tilde{\bar{e}}_\gamma}^2 \sin^2\chi
)/4$ is the sum of the soft scalar masses for $Q$, $\bar{u}$, $L_\alpha$,  $\bar{e}_\beta$, and $\bar{e}_\gamma$. $M_F$ and $M_S$ correspond to the SUSY breaking scale and the messenger scale, respectively. This potential is almost flat for $|\phi|\gg M_S$. $V_{\rm grav}$ is the gravity-mediated contribution, which is always present due to supergravity,
\begin{align}
    V_{\rm grav}=m_{3/2}^2|\phi|^2.
    \label{eq:VSUGRA}
\end{align}
The gravitino mass in the gauge-mediated SUSY breaking scenario is typically smaller than the mass of the AD field, i.e., $m_{3/2}<m_\phi$.

During inflation and the subsequent period of inflaton oscillations, additional SUSY-breaking effects from the inflaton sector lead to the Hubble-induced terms \cite{Dine:1995uk,Dine:1995kz}:\footnote{The thermal effects~\cite{Allahverdi:2000zd,Anisimov:2000wx,Fujii:2001zr} are negligible in the parameter regions of our interest, which is the case for a large field value for $\phi$ and a low reheating temperature.}
\begin{align}
V(\phi) = V_0(\phi) + V_{\rm H}(\phi),\quad
V_{\rm H}(\phi) = -c_1 H^2 |\phi|^2 + c_2 H^2 \frac{|\phi|^4}{\MP^2}+\cdots\,,
\label{eq:Hubble-mass}
\end{align}
where $H$ is the Hubble parameter during and after inflation.
The coefficients $c_1$ and $c_2$ are ${\cal O}(1)$ and depend on the inflation model, and the ellipsis indicates higher-order terms suppressed by the Planck scale.

Generating large lepton flavor asymmetries requires a large field amplitude (see Eq.~\eqref{eq:phi_number}). 
This implies that, during and after inflation, the field value satisfies $|\phi| \gg M_F^2/m_{3/2} > M_S$, so that $V_{\rm grav}$ dominates over $V_{\rm gauge}$.
At that time, the potential is approximately given by
\begin{align}
    V(\phi)\simeq \left(m_{3/2}^2-c_1H^2\right)|\phi|^2 +\left(\frac{\lambda m_{3/2}^2}{4\MP^2}\phi^4+{\rm h.c.}\right)
+c_2 H^2 \frac{|\phi|^4}{\MP^2}+\cdots\,.
\label{eq:Vphi_inf}
\end{align}
We assume $c_1, c_2>0$ and $H>m_{3/2}$ during inflation, and that the AD field acquires a large expectation value $|\phi_0|\lesssim \MP$ during the inflation.
We also assume $\lambda\lesssim 1$ to avoid unwanted charge-breaking minimum at $|\phi_0|\lesssim \MP$.
After the inflation, the AD field remains at this value until the Hubble parameter becomes comparable to the AD field mass, $H = H_{\rm osc} \simeq m_{3/2}$, at which point the AD field begins its coherent oscillation. 
At this time, the AD field also moves in the phase direction due to the lepton-flavor-violating term in Eq.~\eqref{eq:V0}, generating lepton flavor asymmetries.
We assume that it occurs before the reheating ends, i.e., $H_{\rm osc}>\Gamma_{\rm inf}$ with $\Gamma_{\rm inf}$ being the inflaton decay rate, which corresponds to a reheating temperature of $T_R\lesssim 7\times 10^{8}~{\rm GeV}(m_{3/2}/1~{\rm GeV})^{1/2}$.

The time evolution of the $\phi$-number in Eq.~\eqref{eq:phi_number} is given by
\begin{align}
\frac{d}{dt}n_\phi + 3Hn_\phi &= 2{\rm Im}\left(\phi\frac{\partial V}{\partial \phi}\right)
=2{\rm Im}\left(\frac{\lambda m_{3/2}^2}{\MP^2}\phi^4\right).
\end{align}
Therefore, the $\phi$-number shortly after the onset of the AD oscillation at $t= t_{\rm osc}=2/(3H_{\rm osc})$ is estimated as
\begin{align}
n_\phi(t_{\rm osc}) &\simeq 
\frac{4}{3}\delta_{\rm CP}
\frac{\lambda m_{3/2}^2 \MP^2}{H_{\rm osc} }
\left(\frac{|\phi_0|}{\MP}\right)^4\simeq \frac{4}{3}\delta_{\rm CP}\lambda m_{3/2}\MP^2\left(\frac{|\phi_0|}{\MP}\right)^4\,,
\label{eq:n_phi_Qball}
\end{align}
where $\delta_{\rm CP}\lesssim 1$ denotes the effective CP-violating phase, which depends on the initial phase of the AD field during inflation. 

After the onset of oscillations, the AD field initially evolves under the dominance of the gravity-mediation potential, $V_{\rm grav}$, during which Q-balls do not form.\footnote{In general, radiative corrections modify the scalar potential as $V_{\rm grav}(\phi)\simeq m_\phi^2 (1+K \ln({|\phi|^2/M^2}))|\phi|^2$, and the ``new"-type Q-ball forms if $K<0$~\cite{Kasuya:2000sc}. Here, we assume $K>0$, which is realized when the AD field includes the third generation squark and/or the gauginos are light~\cite{Enqvist:2000gq}. 
}
As the field amplitude decreases, the gauge-mediation potential, $V_{\rm gauge}\sim M_F^4$, eventually takes over at $|\phi|=\phi_{\rm eq}\equiv M_F^2/m_{3/2}>M_S$, triggering the formation of so-called delayed-type Q-balls~\cite{Kasuya:2001hg}.
Here and hereafter, we assume $\phi_{\rm eq}<|\phi_0|\lesssim \MP$, which corresponds to $M_F\lesssim 10^9~{\rm GeV}(m_{3/2}/1~{\rm GeV})^{1/2}$.
Numerical simulations have confirmed that the $\phi$-number generated by the AD mechanism --- which, in our case, is proportional to the lepton flavor asymmetries via Eq.~\eqref{eq:nLi_nphi} --- is entirely confined within Q-balls~\cite{Kasuya:1999wu,Kasuya:2001hg}.
Their initial charge is given by~\cite{Kasuya:2001hg}
\begin{align}
    Q_i &\simeq \beta \left(\frac{\phi_{\rm eq}}{M_F}\right)^4
    \notag
    \\
    &\simeq
    6\times 10^{19}
    \left(\frac{m_{3/2}}{1~{\rm GeV}}\right)^{-4}
\left(\frac{M_F}{10^6~{\rm GeV}}\right)^4
\widehat{\beta}
\label{eq:Qinit}
\end{align}
where $\beta\simeq 6\times 10^{-5}$ for $\epsilon\sim \delta_{\rm CP}\lambda (|\phi_0|/\MP)^2\lesssim 0.1$~\cite{Kasuya:2001hg,Kasuya:2012mh} 
and $\widehat{\beta}=\beta/(6\times 10^{-5})$.
The AD field angular velocity, the Q-ball radius, and the Q-ball mass are given by~\cite{Laine:1998rg,Hisano:2001dr}
\begin{align}
\omega_Q &\simeq \sqrt{2}\pi\zeta M_F Q^{-1/4},
\quad
R_Q\simeq \pi \omega_Q^{-1},
\quad
M_Q\simeq \frac{4}{3}\omega_Q Q,
\label{eq:Q-ball-properties}
\end{align}
where $\zeta$ is a numerical coefficient of order unity \cite{Hisano:2001dr,Kawasaki:2012gk}, which is given by $\zeta\simeq 3-4$ in the parameter regions of our interest.
The initial value of $\omega_Q$ is given by 
\begin{align}
\omega_{Q,i}&\simeq 
\sqrt{2}\pi \zeta \beta^{-1/4}m_{3/2}\notag\\
&\simeq
150~{\rm GeV}
\left(\frac{m_{3/2}}{1~{\rm GeV}}\right)
\widehat{\beta}^{-1/4}
\widehat{\zeta},
\label{eq:omega_Q}
\end{align}
where $\widehat{\zeta}=\zeta/3$ and we have used Eq.~\eqref{eq:Qinit}.

At late times, Q-balls gradually release their stored charge through decays into quarks and leptons~\cite{Cohen:1986ct,Kawasaki:2012gk,Kasuya:2012mh}.
The rate of $\phi$-number depletion per unit time is given by~\cite{Kawasaki:2012gk},
\begin{align}
    -\frac{dQ}{dt}\biggl|_{\rm decay}\simeq N\frac{\omega_Q^3}{12\pi^2}4\pi R_Q^2
    \simeq \frac{\pi N}{3}\omega_Q,
    \label{eq:Gamma_Q}
\end{align}
where $N$ denotes the effective number of decay channels contributing to the emission, 
and we have used $R_Q\simeq \pi \omega_Q^{-1}$ in the second equality.
The rate is independent of the coupling due to the Pauli exclusion principle.
In the present scenario, $N$ is given by $N=12$, as shown in Appendix~\ref{app:Ndof}.
Using $dQ/dt\propto Q^{-1/4}$, the Q-ball charge evolves as
\begin{align}
Q(t)\simeq Q_i\left(1-\frac{t}{\tau_Q}\right)^{4/5},
\label{eq:Q_t}
\end{align}
where the Q-ball lifetime $\tau_Q$ is given by
\begin{align}
\tau_Q &\simeq \frac{4}{5}\left(\left.-\frac{1}{Q}\frac{dQ}{dt}\right|_{\rm decay}\right)_{Q=Q_i}^{-1}
\notag\\
&\simeq
1.7\times 10^{-8}{\rm sec}\times
\left(\frac{m_{3/2}}{1~{\rm GeV}}\right)^{-5}
\left(\frac{M_F}{10^6~{\rm GeV}}\right)^{4}
\widehat{\beta}^{5/4}
\widehat{\zeta}^{-1}
\widehat{N}^{-1}
\end{align}
where $\widehat{N}=N/12$.
This corresponds to the Q-ball decay
temperature:
\begin{align}
    T_D&\simeq\left(\frac{\pi^2 g_\ast(T_D)}{90}\right)^{-1/4}\sqrt{\frac{M_P}{2 \tau_Q}}
    \nonumber \\
    &\simeq 4.0~{\rm GeV}~\left(\frac{g_\ast(T_D)}{80}\right)^{-1/4}
    \left(\frac{m_{3/2}}{1~{\rm GeV}}\right)^{5/2}
    \left(\frac{M_F}{10^6~{\rm GeV}}\right)^{-2}
    \widehat{\beta}^{-5/8}
    \widehat{\zeta}^{1/2}
        \widehat{N}^{1/2},
    \label{eq:TD}
\end{align}
where $g_\ast(T_D)$ is the effective relativistic degree of freedom at $T=T_D$.
Depending on $M_F$ and $m_{3/2}$, the Q-ball decay can take place after sphaleron decoupling but before BBN. 
In particular, large lepton flavor asymmetries can be produced before the QCD transition \cite{Asakawa:1989bq,Schwarz:2009ii,Middeldorf-Wygas:2020glx,Vovchenko:2020crk,Gao:2021nwz,Ferreira:2025zeu,Formaggio:2025nde} and the resonant production of sterile neutrino dark matter \cite{Akita:2025txo}.

Let us estimate the resultant lepton flavor asymmetries.
We first show that, in the typical parameter region of interest, Q-balls naturally come to dominate the energy density of the Universe, as can be seen from the ratio of the Q-ball energy density to the radiation energy density at the time of reheating,
\begin{align}
\left. 
\frac{\rho_Q}{\rho_{\rm rad}}
\right|_{T_R}
\simeq
\left. 
\frac{\rho_Q}{\rho_{\rm inf}}
\right|_{t_{\rm osc}}
\simeq \frac{m_{3/2}^2 |\phi_0|^2}{3 \MP^2 H_{\rm osc}^2}
\simeq \frac{1}{3} \left(\frac{|\phi_0|}{\MP}\right)^2,
\label{eq:rho_Q_rho_rad}
\end{align}
where $\rho_Q(t_{\rm osc})\simeq m_{3/2}^2 |\phi_0|^2$ is understood as the Q-ball energy density extrapolated back to $t_{\rm osc}$.
Since $\rho_Q/\rho_{\rm rad}\propto T^{-1}$ in radiation-dominated Universe, the Q-ball energy density comes to dominate the Universe by the time of their decay as long as $T_D\ll T_R(|\phi_0|/\MP)^2$. The resulting lepton flavor asymmetries are then given by 
 \begin{align}
 Y_{L_\alpha}
= 
\left.\frac{\Delta n_{L_\alpha}}{s}\right|_{T_D}
&\simeq 
\left.
\frac{\rho_Q}{s}
\right|_{T_D}
\left. 
\frac{\Delta n_{L_\alpha}}{\rho_Q}
\right|_{t_{\rm osc}}
\nonumber \\
    &\simeq 
    \frac{\kappa_\alpha\delta_{\rm CP}\lambda}{4}\frac{T_D}{m_{3/2}}\left(\frac{|\phi_0|}{\MP}\right)^2, \nonumber \\
    &\simeq 
    0.10 \kappa_\alpha
    \left(\frac{\delta_{\rm CP} \lambda}{0.1}\right)
    \left(\frac{T_D}{4~{\rm GeV}}\right)
    \left(\frac{m_{3/2}}{1~{\rm GeV}}\right)^{-1}
    \left(\frac{|\phi_0|}{\MP}\right)^2,
    \label{eq:YL_withQ_1}
\end{align}
where we have used 
$\Delta n_{L_\alpha}/\rho_Q|_{T_D}=\Delta n_{L_\alpha}/\rho_Q|_{t_{\rm osc}}$,
$\rho_Q/s(T_D)\simeq (3/4)T_D$,
$\rho_Q(t_{\rm osc})\simeq m_{3/2}^2 |\phi_0|^2$,
Eqs.~\eqref{eq:nLi_nphi}, \eqref{eq:n_phi_Qball}
and neglected corrections of $\lesssim \mathcal{O}(1)$ from the finite chemical potentials. 
Using Eq.~\eqref{eq:TD}, it can also be written as
 \begin{align}
 Y_{L_\alpha}
    &\simeq 
    0.10 \kappa_\alpha
    \left(\frac{\delta_{\rm CP} \lambda}{0.1}\right)
    \left(\frac{g_*(T_D)}{80}\right)^{-1/4}    
    \left(\frac{m_{3/2}}{1~{\rm GeV}}\right)^{3/2}
    \left(\frac{M_F}{10^6~{\rm GeV}}\right)^{-2}    
    \left(\frac{|\phi_0|}{\MP}\right)^2
    \widehat{\beta}^{-5/8}
    \widehat{\zeta}^{1/2}
    \widehat{N}^{1/2}.
    \label{eq:YL_withQ_2}
\end{align}
Therefore, depending on the flat direction and the angle $\chi$, large lepton flavor asymmetries with an arbitrary ratio of $Y_{L_e}$, $Y_{L_\mu}$, and $Y_{L_\tau}$ can be generated while maintaining the vanishing total lepton number $Y_L=0$ without fine-tuning.

\subsection{Small baryon asymmetry from partial sphaleron conversion}
\label{subsec:BAU_from_partial_conversion}

Lepton flavor asymmetries emitted from Q-balls before the electroweak phase transition are converted to baryon asymmetry via the sphaleron process. Three processes can be considered in this context: decay~\cite{Cohen:1986ct,Kawasaki:2012gk,Kasuya:2012mh}, and evaporation and diffusion~\cite{Laine:1998rg,Banerjee:2000mb}. 
However, as we show in Appendix~\ref{app:evap-diff}, the evaporation and the diffusion give negligible contributions to the baryon asymmetry in most of the parameter regions of our interest. Therefore, we focus on the contribution of the decay in this section.

From Eq.~\eqref{eq:Gamma_Q}, the Q-ball charge emitted before the sphaleron decoupling is given by
\begin{align}
\Delta Q\simeq 
\int dt\left|\frac{dQ}{dt}\right|
\simeq \frac{\pi N}{3}\omega_Q t_{\rm sph},
\label{eq:DeltaQ_decay}
\end{align}
where $t_{\rm sph}=2/3H_{\rm sph}$ is the cosmic time of the sphaleron decoupling. $H_{\rm sph}$ is related to the temperature $T_{\rm sph}\simeq 130~{\rm GeV}$, depending on the dominant source of the radiation at that time.
There are two possible sources: the radiation from the inflaton decay, i.e., reheating, and that from the Q-ball decay~\cite{Kawasaki:2002hq}. After the reheating, the former component dominates the radiation for a while. During this period, we have
\begin{align}
\left.\frac{\rho_Q}{s}\right|_{T}
=\left.\frac{\rho_Q}{s}\right|_{T_R}
=\frac{1}{4}T_R\left(\frac{|\phi_0|}{\MP}\right)^2
=\frac{1}{4}T_{R,{\rm eff}},
\label{eq:TReff}
\end{align}
where we have used Eq.~\eqref{eq:rho_Q_rho_rad} and $\rho_{\rm rad}/s|_{T_R} = (3/4)T_R$, and defined $T_{R,{\rm eff}}=T_R (|\phi_0|/\MP)^2$ for notational simplicity.
Using $\rho_Q=3\MP^2 H^2$ and $s=(2\pi^2 g_*/45)T^3$, we have
\begin{align}
T = 
\left(\frac{\pi^2 g_*(T)}{270}\right)^{-1/3}
\left(\frac{H^2 \MP^2}{T_{R,{\rm eff}}}\right)^{1/3}.
\label{eq:T_from_inf}
\end{align}
On the other hand, the radiation from the Q-ball decay dominates at late time, whose temperature is given by~\cite{Kolb:1990vq}
\begin{align}
T\simeq (T_D^2 \MP H)^{1/4}.
\label{eq:T_from_Q}
\end{align}
The transition from the inflaton-induced radiation to the Q-ball-induced radiation occurs at
\begin{align}
T_p &\simeq (T_D^4 T_{R,{\rm eff}})^{1/5} 
\left(\frac{\pi^2 g_*(T_p)}{270}\right)^{1/5}
\notag\\
&\simeq 39~{\rm GeV}\
\left(\frac{g_*(T_p)}{100}\right)^{1/5}
\left(\frac{T_D}{4~{\rm GeV}}\right)^{4/5}
\left(\frac{T_R}{10^5~{\rm GeV}}\right)^{1/5}
\left(\frac{|\phi_0|}{\MP}\right)^{2/5}.
\label{eq:Tp}
\end{align}
The relation between the temperature $T$ and the Hubble expansion rate $H$ is given by Eq.~\eqref{eq:T_from_inf} for $T>T_p$ and Eq.~\eqref{eq:T_from_Q} for $T<T_p$.
Therefore, Eq.~\eqref{eq:DeltaQ_decay} results in
\begin{align}
    \Delta Q &\simeq 
\frac{2\pi N \omega_Q}{9}\cdot 
\begin{cases}
\displaystyle{
\left(\frac{\pi^2 g_*(T_{\rm sph})}{270}\right)^{-1/2}
\frac{M_P}{(T_{R,{\rm eff}} T_{\rm sph}^3)^{1/2}}
}
& (T_p<T_{\rm sph})
\\
\displaystyle{
\frac{T_D^2 \MP}{T_{\rm sph}^4}
}
& (T_p>T_{\rm sph})
\end{cases}.
\end{align}
Using Eqs.~\eqref{eq:Qinit}, \eqref{eq:omega_Q} and \eqref{eq:TD}, we obtain
\begin{align}
    \frac{\Delta Q}{Q_i} &\simeq 
    \begin{cases}
    \displaystyle{
    5.7\times 10^{-5} 
    \left(\frac{m_{3/2}}{1~{\rm GeV}}\right)^{5}
    \left(\frac{M_F}{10^6~{\rm GeV}}\right)^{-4}
    \left(\frac{T_R}{10^5~{\rm GeV}}\right)^{-1/2}    
    \left(\frac{|\phi_0|}{\MP}\right)^{-1}    
    \widehat{\beta}^{-5/4}
    \widehat{\zeta}
    \widehat{N}
    }
    & (T_p < T_{\rm sph})
    \\
    \displaystyle{
    2.9\times 10^{-6}
    \left(\frac{m_{3/2}}{1~{\rm GeV}}\right)^{10}
    \left(\frac{M_F}{10^6~{\rm GeV}}\right)^{-8}  
    \left(\frac{g_*(T_D)}{80}\right)^{-1/2}      
    \widehat{\beta}^{-5/2}
    \widehat{\zeta}^2
    \widehat{N}^2
    }    
    & (T_p > T_{\rm sph})
    \end{cases}
\end{align}
where we have used $T_{\rm sph}\simeq 130~{\rm GeV}$ and $g_*(T_{\rm sph})\simeq 100$.

The lepton flavor asymmetries emitted before the electroweak phase transition, evaluated at $T=T_D$, is given by
\begin{align}
\Delta Y_{L_\alpha}
\simeq \frac{\Delta Q}{Q_i}Y_{L_\alpha},
\label{eq:YLa_before_EWPT}
\end{align}
where $Y_{L_\alpha}$ is given by Eq.~\eqref{eq:YL_withQ_1} or Eq.~\eqref{eq:YL_withQ_2}.
These asymmetries are partially converted into the baryon asymmetry via the sphaleron process~\cite{Kuzmin:1985mm}.
For a total zero lepton asymmetry, the conversion cancels out, but not completely, due to differences in the lepton Yukawa couplings~\cite{Khlebnikov:1988sr,March-Russell:1999hpw,Laine:1999wv}. The resultant baryon asymmetry is given by~\cite{Khlebnikov:1988sr,March-Russell:1999hpw,Laine:1999wv,Mukaida:2021sgv}
\begin{align}
    Y_B=
\frac{\Delta n_B}{s}\simeq-0.030
\left(h_{\tau}^2 \Delta Y_{L_\tau}+h_{\mu}^2 \Delta Y_{L_\mu}+h_e^2 \Delta Y_{L_{e}} \right),
\label{eq:YB_vs_YLa}
\end{align}
where $h_\tau\simeq 0.010$, $h_\mu\simeq 6.1\times 10^{-4}$, and $h_e\simeq 2.9\times 10^{-6}$ are the SM lepton Yukawa couplings, and
$\Delta Y_{L_\alpha}$ is given by Eq.~\eqref{eq:YLa_before_EWPT}. 
The baryon asymmetry is therefore doubly suppressed: by the partial Q-ball charge emission $\Delta Q/Q_i$, and by the near cancellation of the sphaleron conversion due to $\sum_{\alpha} Y_{L_\alpha}=0$. 
For example, for $|Y_{L_\tau}|\gtrsim 0.004\ |Y_{L_\mu}|$, we have
\begin{align}
Y_B\simeq 3\times 10^{-10}
\left(\frac{\Delta Q/Q_i}{10^{-3}}\right)
\left(\frac{Y_{L_\tau}}{-0.1}\right),
\end{align}
which can naturally explain the small observed baryon asymmetry in the typical parameter region considered in this work.

\begin{figure}[!t]
\centering
\begin{tabular}{cc}
\hspace{-0.075cm}\includegraphics[width=0.5\textwidth]{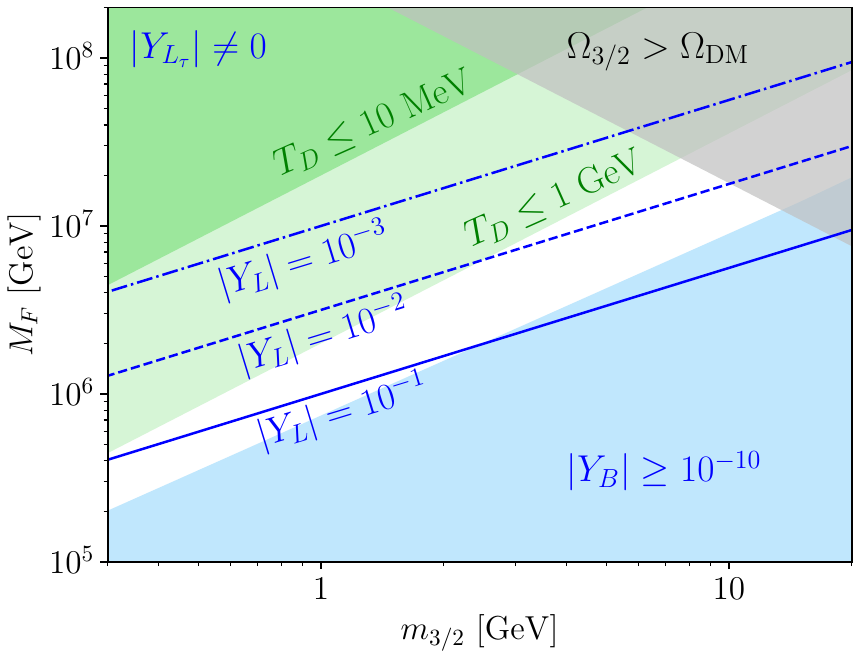}
\hspace{-0.075cm}\includegraphics[width=0.5\textwidth]{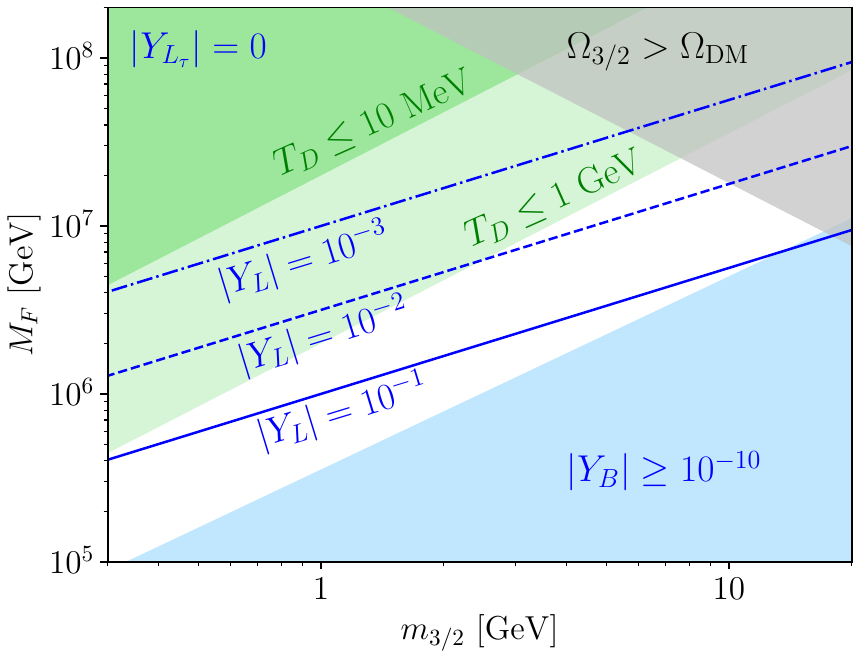}
\end{tabular}
\vspace{-0.2cm}
\caption{Typical magnitude of lepton flavor asymmetries, $|Y_L|$, with zero total lepton asymmetry in the Affleck-Dine leptoflavorgenesis. The left (right) panel shows the case with (without) tau flavor asymmetry since the resultant baryon asymmetry depends on the existence of tau flavor asymmetry (cf. Eq.~\eqref{eq:YB_vs_YLa}).
We set the flavor ratio $|\kappa_\tau|=1$ with $\sum_\alpha \kappa_\alpha=0$ and $Y_L=Y_{L_\tau}$ (left), 
$\kappa_e=-\kappa_\mu=1,~ \kappa_\tau=0$ and $Y_L=Y_{L_\mu}$ (right), $m_\phi=10^4~{\rm GeV}$, $\delta_{\rm CP}\lambda=0.1,~|\phi_0|= \MP$, $\widehat{\beta}=\widehat{\zeta}=\widehat{N}=1$ and the reheating temperature $T_R=10^5~{\rm GeV}$ (cf.~Appendix~\ref{app:list_of_paras}).
The blue solid, dashed, dot-dashed lines correspond to $Y_L=10^{-1},~10^{-2}$ and $10^{-3}$, respectively.
In the light green region, the Q-ball decay temperature is $T_D\leq1~{\rm GeV}$. For $T_D\gtrsim1~{\rm GeV}$, lepton flavor asymmetries would be generated early enough to affect the QCD transition~\cite{Asakawa:1989bq,Schwarz:2009ii,Middeldorf-Wygas:2020glx,Vovchenko:2020crk,Gao:2021nwz,Ferreira:2025zeu,Formaggio:2025nde} and the production of sterile neutrino dark matter~\cite{Akita:2025txo}. 
In the dark green region, $T_D\lesssim 10~{\rm MeV}$, lepton asymmetries are generated after the onset of neutrino oscillations~\cite{Domcke:2025lzg}, which may not be sufficiently washed out and may be inconsistent with the BBN observation.
In the light blue region, the baryon asymmetry is overproduced, $|Y_B|\geq10^{-10}$. On the contour of the light blue region, the observed baryon asymmetry can be explained if the tau (mu) flavor asymmetry has a sign opposite to that of the observed baryon asymmetry in the left (right) panels (cf. Eq.~\eqref{eq:YB_vs_YLa}).
In the gray region, gravitinos are overproduced compared to the observed dark matter abundance, $\Omega_{\rm 3/2}>\Omega_{\rm DM}$.
}
\label{fig:parameterspace_L}
\end{figure}

\subsection{Results}

Figure~\ref{fig:parameterspace_L} shows typical magnitudes of lepton flavor asymmetries with zero total lepton asymmetry. We consider two cases with (without) tau flavor asymmetry in the left (right) panels since the resultant baryon asymmetry significantly depends on tau flavor asymmetry. 
We set $|\kappa_\alpha|=1$ with $\sum_\alpha \kappa_\alpha=0$ and $Y_L=Y_{L_\tau}$
in the left panels while we set $\kappa_e=-\kappa_\mu=1,~\kappa_\tau=0$ and $Y_L=Y_{L_\mu}$ in the right panels. 
In both panels, we use $m_\phi=10^4~{\rm GeV}$, $\delta_{\rm CP}\lambda=0.1$, $|\phi_0|= \MP$,
$\widehat{\beta}=\widehat{\zeta}=\widehat{N}=1$ and the reheating temperature $T_R=10^5~{\rm GeV}$.
The blue solid, dashed, dot-dashed lines correspond to $Y_L=10^{-1},~10^{-2}$ and $10^{-3}$, respectively.
The light green region shows $T_D\leq1~{\rm GeV}$. For $T_D\gtrsim1~{\rm GeV}$, lepton flavor asymmetries would be generated early enough to affect the QCD transition and the production of sterile neutrino dark matter~\cite{Akita:2025txo}. 
In the dark green region, $T_D\lesssim 10~{\rm MeV}$, lepton asymmetries are generated after the onset of neutrino oscillations~\cite{Domcke:2025lzg}, which may not be sufficiently washed out and may be inconsistent with the BBN observation.
In the light blue region, the baryon asymmetry is overproduced, $Y_B\geq10^{-10}$. In the gray region, gravitinos are overproduced compared to the observed dark matter abundance, $\Omega_{\rm 3/2}>\Omega_{\rm DM}$.

In the contour of the light blue region, the observed baryon asymmetry of $Y_B\sim 10^{-10}$ is generated if the tau (mu) flavor asymmetry has a sign opposite to that of the observed baryon asymmetry in the left (right) panels. The observed baryon asymmetry can be explained in the region of large lepton flavor asymmetries such as $Y_{L_\tau}\sim-10^{-1}\cdots-10^{-2}$ or $Y_{L_\mu}\sim-10^{-1}$ with $Y_{L_\tau}\simeq0$.

In the dark green region of $T_D\leq 10~{\rm MeV}$, lepton flavor asymmetries are generated after the onset of flavor-changing neutrino oscillations. The final asymmetries after neutrino decoupling and their impacts on subsequent cosmology, such as BBN, are nontrivial. On the other hand, if the asymmetries are generated at $T\gtrsim 10~{\rm MeV}$, we can use the results of Ref.~\cite{Domcke:2025lzg}, where the authors have revealed the evolutions of lepton flavor asymmetries for their initial values at $T\gtrsim 10~{\rm MeV}$ before neutrino oscillations start.

Lastly, we will discuss constraints on our scenario from other physical phenomena.
First, we discuss the gravitino problem. In the gauge-mediated SUSY breaking scenario, gravitinos are the stable lightest supersymmetric particles.
They are mostly produced at $T\sim T_R$~\cite{Moroi:1993mb}, and their abundance is given by~\cite{Bolz:2000fu,Pradler:2006qh,Rychkov:2007uq,Eberl:2020fml}
\begin{align}
\Omega_{3/2}h^2 &\simeq 0.3
\left(\frac{m_{3/2}}{1~{\rm GeV}}\right)^{-1}
\left(\frac{T_R}{10^5~{\rm GeV}}\right)
\left(\frac{M_{\tilde{g}}}{10^4~{\rm GeV}}\right)^2
\left(\frac{\widehat{\gamma}_{3/2}}{0.4}\right)
\times
\left.\frac{s_{\rm before}}{s_{\rm after}}\right|_{T_D}
\notag\\
&\simeq
4\times 10^{-5}
\left(\frac{m_{3/2}}{1~{\rm GeV}}\right)^{-1}
\left(\frac{T_D}{4~{\rm GeV}}\right)
\left(\frac{|\phi_0|}{\MP}\right)^{-2}
\left(\frac{M_{\tilde{g}}}{10^4~{\rm GeV}}\right)^2
\left(\frac{\widehat{\gamma}_{3/2}}{0.4}\right),
\end{align}
where $M_{\tilde{g}}$ is the gluino mass, $\widehat{\gamma}_{3/2}\lesssim 0.4$ is a numerical factor~\cite{Rychkov:2007uq,Eberl:2020fml}, and $h\sim 0.7$ is the scaling factor for the Hubble constant~\cite{ParticleDataGroup:2024cfk}.
Here, we have included a dilution factor by the Q-ball decay, $s_{\rm before}/s_{\rm after}|_{T_D}=3T_D/T_{R,{\rm eff}}$.
Therefore, in the parameter regions in Fig.~\ref{fig:parameterspace_L}, the gravitino abundance from the reheating is much smaller than the observed dark matter abundance, $\Omega_{\rm DM}h^2\simeq 0.12$~\cite{ParticleDataGroup:2024cfk}.

Gravitinos can also be directly produced from Q-ball decay. As shown in Appendix~\ref{app:Gravitino_from_Q}, their abundance is given by
\begin{align}
\Omega_{3/2}h^2
\simeq 1.1\times 10^{-6}
\left(\frac{m_{3/2}}{1~{\rm GeV}}\right)^{5/2}
\left(\frac{M_F}{10^6~{\rm GeV}}\right)^2
\left(\frac{M_{\tilde{g}}}{10^4~{\rm GeV}}\right)^{-2}
\left(\frac{\delta_{\rm CP}\lambda}{0.1}\right)
\left(\frac{|\phi_0|}{\MP}\right)^2
\widehat{\beta}^{-9/8}
\widehat{\zeta}^{13/2}
\widehat{N}^{1/2}.
\end{align}
This should be smaller than $\Omega_{\rm DM}h^2\simeq 0.12$~\cite{ParticleDataGroup:2024cfk}, which excludes the top-right corner of Fig.~\ref{fig:parameterspace_L}.

We also comment on constraints from the chiral plasma instability \cite{Domcke:2022uue}. Lepton asymmetries can induce a helical hypermagnetic field through the chiral plasma instability, which can subsequently generate the baryon asymmetry of the Universe. 
To avoid the overproduction of the baryon asymmetry, lepton flavor asymmetries are constrained at a temperature slightly above the electron Yukawa equilibration.  
The constraint includes large theoretical uncertainties (see also Ref.~\cite{Hamada:2025cwu}) and is not straightforwardly applicable to the current scenario, since the evolution of hypermagnetic fields and their conversion into baryon asymmetry can be modified under Q-ball domination. In addition, Q-balls protect lepton flavor asymmetries from the chiral plasma instability. 

Here, we give a rough estimate focusing on the benchmark point which will be introduced in Sec.~\ref{sec:implication}, and assuming $T_R\sim T_{R,{\rm eff}}\simeq 10^5~{\rm GeV}$, for simplicity. In that case, the electron Yukawa equilibration occurs at $T_e\sim 10^5~{\rm GeV}\left(T_R/10^5~{\rm GeV}\right)^{-1}$, as in the radiation dominated Universe. Following Ref.~\cite{Domcke:2022uue}, we evaluate the lepton asymmetry at a temperature $T_e'\sim 10T_e\sim 10^6~{\rm GeV}$. We assume that the chemical potential is bounded from above as $\mu_\alpha\lesssim \omega _Q$ at this stage, by the chemical equilibration of the Q-ball and the plasma, see Appendix~\ref{app:evap-diff}. Using $\omega_Q\simeq 1100~{\rm GeV}$ for the benchmark point, we have $\xi_\alpha \sim\mu_\alpha/T'_e\lesssim \omega_Q/T'_e\sim 10^{-3}$. This asymmetry is small enough to avoid the chiral plasma instability~\cite{Domcke:2022uue,Kamada:2018tcs}. Therefore, we expect that the constraint can be avoided at this benchmark point. A more detailed analysis of the effect of the chiral plasma instability in the presence of Q-balls is beyond the scope of this work and is left for future study.


\section{Implications for early Universe cosmology}
\label{sec:implication}

In this section, we consider a benchmark point for large lepton flavor asymmetries and briefly review how such a benchmark could simultaneously affect various aspects of early Universe cosmology: (i) explaining the observed baryon asymmetry by the same origin as the lepton flavor asymmetries, (ii) affecting the nature of the QCD transition, (iii) opening up a new parameter space of sterile neutrino dark matter by enhancing their production and (iv) altering the abundance of the light elements, in particular, resolving the recently reported helium-4 anomaly.
We show that the ADLFG scenarios can generate such a benchmark point.
In this benchmark point, we assume normal hierarchy (NH) of neutrino masses. We will briefly comment on the case of their inverted hierarchy (IH) in the last paragraph of this section.

We consider the following benchmark point, which can be realized in the ADLFG 
discussed in Sec.~\ref{sec:ADLFG_with_Q},
\begin{align}
    Y_{L_e} = 0.06,\ \ \ \ Y_{L_\mu}= -0.03,\ \ \ \  Y_{L_\tau} = -0.03\, . 
    \label{benchmark_Qball}
\end{align}
First, let us see the BBN and CMB constraints, and the related helium-4 anomaly.
At temperatures well below the QCD transition, only leptons and photons remain abundant in the plasma, and lepton asymmetries are stored in neutrinos due to the charge neutrality of the plasma.
Lepton flavor asymmetries, $Y_{L_\alpha}=(n_{L_\alpha}-n_{\bar{L}_{\alpha}})/s$, are related to neutrino chemical potentials, $\xi_\alpha=\mu_\alpha/T$, as
\begin{align}
    n_{L_\alpha} -n_{\bar{L}_{\alpha}} &=\frac{T^3}{6}\left(\xi_\alpha + \frac{\xi_\alpha^3}{\pi^2} \right), \\
    Y_{L_\alpha}&\equiv \frac{n_{L_\alpha}-n_{\bar{L}_{\alpha}}}{s}\simeq\frac{1/6}{2\pi^2 g_\ast/45}\left(\xi_\alpha + \frac{\xi_\alpha^3}{\pi^2}\right)\simeq 0.035\xi_\alpha\left(1+\frac{\xi^2_\alpha}{\pi^2}\right),
    \label{eq:YL_vs_xi}
\end{align}
with $g_\ast=10.75$ as the value of the BBN epoch.
Lepton flavor asymmetries are conserved until neutrino oscillations start at $T\sim 15~{\rm MeV}$.
The benchmark point corresponds to the following chemical potentials at $T\sim 15~{\rm MeV}$,
\begin{align}
    \xi_e^{\rm ini}\simeq 1.4 ,\ \ \ \ \xi_\mu^{\rm ini}\simeq-0.8,\ \ \ \ \xi_\tau^{\rm ini} \simeq-0.8\, . 
\end{align}
Below $T\sim 15~{\rm MeV}$, to track the evolution of lepton flavor asymmetries, we need to solve the kinetic equations for neutrinos and the plasma, including neutrino oscillations, until neutrino decoupling.
We assume that for the benchmark point~\eqref{benchmark_Qball}, lepton flavor asymmetries are generated at $T\gg 15~{\rm MeV}$ before neutrino oscillations start. Then, we can use the results of Ref.~\cite{Domcke:2025lzg}, where the authors comprehensively study the evolution of lepton flavor asymmetries at $T\lesssim 15~{\rm MeV}$ by solving the momentum-averaged quantum kinetic equations (QKEs) for neutrinos. 

The light elements, ${\rm D,~^3He,~^4He}$ and ${\rm ^7Li}$ would start to be synthesized at $T\simeq 1~{\rm MeV}$.
Their abundances are sensitive to electron flavor asymmetry through the freeze-out of the neutron-proton ratio. The electon flavor chemical potential for the benchmark point at $T=1~{\rm MeV}$ is, assuming normal hierarchy of neutrino masses, (see Fig.~1 in Ref.~\cite{Domcke:2025lzg}),
\begin{align}
    \xi_e^{\rm BBN}\simeq 0.04.
\end{align}
This value is favored by the recent EMPRESS measurement of the helium-4 abundance \cite{Matsumoto:2022tlr,Escudero:2022okz,Froustey:2024mgf,Domcke:2025lzg,Burns:2022hkq,Yanagisawa:2025mgx}.

The negative tau flavor asymmetry produces a small and positive baryon asymmetry through the sphaleron transition, suppressed by the Q-ball protection. We will show that this benchmark point can generate the observed baryon asymmetry in our scenario later.

Furthermore, large lepton flavor asymmetries may lead to a first-order QCD phase transition~\cite{Gao:2021nwz,Gao:2024fhm}. This requires the asymmetries of $|Y_{L_\alpha}|\sim10^{-3}\cdots10^{-1}$, depending on the direction in flavor space. However, there may still remain a lot of uncertainties. In addition, the QCD phase transition at this benchmark point has not yet been studied explicitly. We leave detailed discussions on the QCD transition for future work. 

Large lepton flavor asymmetries also enhance the production of sterile neutrino dark matter by inducing the Mikheyev-Smirnov-Wolfenstein (MSW)-like resonance. Our previous work~\cite{Akita:2025txo} shows that the asymmetries of $|Y_{L_\alpha}|\sim 10^{-3}\cdots10^{-1}$ open up a new parameter space of sterile neutrino dark matter. We have also confirmed that at this benchmark point, sterile neutrinos can constitute all dark matter, without confronting observational constraints.\footnote{Our code \texttt{sterile-dm-lfa} in Ref.~\cite{Akita:2025txo} is publicly available on \gitlink~, which traces the evolution of sterile neutrinos in the presence of large lepton flavor asymmetries.}

Notably, the ADLFG scenario with Q-balls can produce these required asymmetries of $|Y_{L_\alpha}|\sim 10^{-3}\cdots10^{-1}$ in any flavor space at $T\gtrsim 1~{\rm GeV}$ as shown in Figure~\ref{fig:parameterspace_L}, which is early enough to affect the QCD transition and the production of sterile neutrino dark matter. 

Finally, we show the ADLFG scenario with Q-ball can successfully produce this benchmark point for lepton flavor asymmetries and the observed positive baryon asymmetry. For this benchmark point of Eq.~\eqref{benchmark_Qball}, the parameters of $\kappa_\alpha$ are fixed as
\begin{align}
    (\kappa_e,\ \kappa_\mu,\ \kappa_\tau) = \left(1,\ -\frac{1}{2},\ -\frac{1}{2} \right).
\end{align}
As a benchmark point for the ADLFG, for example, let us take $m_{3/2}\simeq 7~{\rm GeV}$ and $M_F\simeq 5.6\times 10^6~{\rm GeV}$. Then,
the temperature of the Q-ball decay, i.e., the temperature at which the asymmetries are generated, and the resultant lepton flavor asymmetries are given by, following Sec.~\ref{sec:ADLFG_with_Q},
\begin{align}
    T_D&\simeq 17~{\rm GeV}~\left(\frac{g_\ast(T_D)}{80}\right)^{-1/4}\left(\frac{m_{3/2}}{7~{\rm GeV}}\right)^{5/2}\left(\frac{M_F}{5.6\times10^6~{\rm GeV}}\right)^{-2}
    \widehat{\beta}^{-5/8}
    \widehat{\zeta}^{1/2}
    \widehat{N}^{1/2}, \\
    Y_{L_e}&\simeq 0.06~\left(\frac{\kappa_e}{1} \right)\left(\frac{\delta_{\rm CP} \lambda}{0.1} \right)\left(\frac{T_D}{17~{\rm GeV}}\right)\left(\frac{m_{3/2}}{7~{\rm GeV}}\right)^{-1}\left(\frac{|\phi_0|}{\MP}\right)^2.
    \\
    Y_{L_\mu}=Y_{L_\tau}&\simeq -0.03~\left(\frac{\kappa_\tau}{-1/2} \right)\left(\frac{\delta_{\rm CP} \lambda}{0.1} \right)\left(\frac{T_D}{17~{\rm GeV}}\right)\left(\frac{m_{3/2}}{7~{\rm GeV}}\right)^{-1}\left(\frac{|\phi_0|}{\MP}\right)^2.
\end{align}
The fraction of the Q-ball decay before the electroweak phase transition, $\Delta Q/Q_i$, and the final baryon asymmetry, $Y_B$, are given by
\begin{align}
    \frac{\Delta Q}{Q_i}&\simeq 9.8\times 10^{-4}
    \left(\frac{m_{3/2}}{7~{\rm GeV}}\right)^{5}
    \left(\frac{M_F}{5.6\times 10^6~{\rm GeV}}\right)^{-4}  \left(\frac{T_R}{10^5~{\rm GeV}}\right)^{-1/2}    
    \left(\frac{|\phi_0|}{\MP}\right)^{-1}    
    \widehat{\beta}^{-5/4}
    \widehat{\zeta}
    \widehat{N},  \\
    Y_B&\simeq 8.8\times 10^{-11}~\left(\frac{\Delta Q/Q_i}{9.8\times10^{-4}}\right)\left(\frac{Y_{L_\tau}}{-0.030}\right),
\end{align}
where we have used $T_p\lesssim T_{\rm sph}$ at this benchmark point with $T_R= 10^5~{\rm GeV}$.
Thus, at this benchmark point, the ADLFG scenario can simultaneously generate the observed baryon asymmetry, resolve the helium-4 anomaly, enhance the production of sterile neutrino dark matter, and may induce the first-order QCD phase transition.

Let us comment on the case of the inverted hierarchy of neutrino masses since we have assumed their normal hierarchy until now. The ADLFG scenario, the nature of the QCD transition, and the production of sterile neutrino dark matter do not depend on the hierarchy of neutrino masses. On the other hand, the nucleosynthesis of the light elements depends on their hierarchy through the washout of lepton flavor asymmetries by neutrino oscillations at $T\lesssim 15~{\rm MeV}$. In particular, $\xi_e^{\rm BBN}\simeq 0.04$ at the BBN epoch, which is favored by the helium-4 anomaly, requires an initial positive tau-flavor asymmetry in the IH case, $\xi^{\rm ini}_\tau>0$, before the onset of neutrino oscillations (see the right panel of Fig.~1 in Ref.~\cite{Domcke:2025lzg}). However, in the ADLFG scenario, the observed baryon asymmetry can be explained by a negative tau-flavor asymmetry (cf.~Eq.~\eqref{eq:YB_vs_YLa}). Thus, in the IH case, the AFDLFG scenario could have broad impacts on cosmology without simultaneously resolving the observed baryon asymmetry and the helium-4 anomaly.

\section{Conclusions}
\label{sec:conclusions}

Large lepton flavor asymmetries with zero total lepton asymmetry would induce rich phenomenology spanning from the early Universe to the current Universe, as outlined in the Introduction. 
They are much less constrained by the BBN and CMB observations, since flavor-violating neutrino oscillations wash out flavor asymmetries at $T\lesssim 15$ MeV, as comprehensively studied in Ref.~\cite{Domcke:2025lzg}. 
Motivated by this, we have proposed an Affleck-Dine leptoflavorgenesis scenario that can generate large lepton flavor asymmetries.
The mechanism allows for the generation of lepton flavor asymmetries in arbitrary directions in flavor space.

In Sec.~\ref{sec:ADLFG_with_Q}, we presented a concrete realization of Affleck-Dine leptoflavorgenesis based on the $Q\bar{u}L\bar{e}$ flat directions.
In this scenario, lepton flavor asymmetries are confined inside Q-balls and protected from the sphaleron transition.
The typical magnitude of lepton flavor asymmetries is $|Y_{L_\alpha}|\sim 10^{-3}\cdots10^{-1}$ as shown in Figure~\ref{fig:parameterspace_L}.
The baryon asymmetry is generated through partial emissions of lepton flavor asymmetries from Q-balls before the sphaleron decoupling.
If tau flavor asymmetry of $Y_{L_\tau}\sim-10^{-2}\cdots-10^{-1}$ or muon flavor asymmetry of $Y_{L_\mu}\sim -10^{-1}$ is produced, the observed baryon asymmetry can be generated by the same origin as the lepton flavor asymmetries in this scenario. 

In Sec.~\ref{sec:implication}, we discussed implications of the Affleck-Dine leptoflavorgenesis for early Universe cosmology.
We highlighted that it could simultaneously have broad impacts on the cosmological puzzles, such as the observed small baryon asymmetry, the nature of the QCD transition, the production of sterile neutrino dark matter, and the abundance of the light elements, in particular, the recently reported helium-4 anomaly.

In summary, we have proposed the Affleck-Dine leptoflavorgenesis scenario, generating large lepton flavor asymmetries summing up to zero total lepton asymmetry with a wide range of magnitudes and in any direction of flavor space. The presented scenarios could have broad impacts on cosmology.

\bigskip

\begin{center}
\textbf{Acknowledgements} 
\end{center}
This work has received support from JSPS Grant-in-Aid
for Scientific Research KAKENHI Grant No.~24KJ0060, No.~24H02244, and No.~24K07041, and from the European Union's Horizon Europe research and innovation programme under the Marie Sklodowska-Curie grant agreement No~101204216.

\appendix

\section{List of parameters}
\label{app:list_of_paras}

Input parameters and key derived quantities used in our analysis are summarized in Table~\ref{tab:list_of_paras}.
\begin{table}[t!]
\begin{tabular}{c|l|c}
\hline
\multicolumn{3}{c}{\textbf{Input parameters}} \\ \hline
Symbol & Description & Eqs.
\\ \hline
$\alpha$ & Lepton flavor index ($e,\mu,\tau$) & --
\\
$\chi$ & Relative angle of the AD flat direction in flavor basis & \eqref{eq:QuLe}
\\
$\lambda$ & Coefficient in the K\"ahler potential & \eqref{eq:V0}
\\
$m_{3/2}$ & Gravitino mass & \eqref{eq:V0},\eqref{eq:VSUGRA}
\\
$m_\phi$ & Soft scalar mass for the AD field & \eqref{eq:Vgauge}
\\
$M_F$ & Scale corresponding to SUSY breaking & \eqref{eq:Vgauge}
\\
$\phi_0$ & Expectation value of the AD field during inflation & below~\eqref{eq:Vphi_inf}
\\
$\delta_{\rm CP}$ & Effective CP violating phase & \eqref{eq:n_phi_Qball}
\\
$\beta$ & Q-ball charge parameter ($\simeq 6\times 10^{-5}$) & \eqref{eq:Qinit}
\\
$\zeta$ & Q-ball properties parameter ($\simeq 3-4$) & \eqref{eq:Q-ball-properties}
\\
$N$ & effective number of Q-ball decay channels ($=12$) & \eqref{eq:Gamma_Q}, App.~\ref{app:Ndof}
\\
$T_{\rm sph}$ & sphaleron decoupling temperature ($\simeq 130~{\rm GeV}$) & --
\\
$T_R$ & reheating temperature after the inflation & --
\\ \hline
\multicolumn{3}{c}{\textbf{Derived quantities}} \\ \hline
Symbol & Description & Eqs.
\\ \hline
$\kappa_\alpha$ & fraction of the lepton flavor asymmetry relative to the $\phi$-number & \eqref{eq:nLi_nphi},\eqref{eq:kappa_and_chi}
\\
$Q_i$ & Initial Q-ball charge & \eqref{eq:Qinit}
\\
$\omega_Q$ & AD field angular velocity & \eqref{eq:Q-ball-properties},\eqref{eq:omega_Q}
\\
$T_D$ & Decay temperature of Q-ball & \eqref{eq:TD}
\\
$Y_{L_\alpha}$ & lepton flavor asymmetry & \eqref{eq:YL_withQ_1},\eqref{eq:YL_withQ_2}
\\
$T_{R,{\rm eff}}$ & $=T_R(|\phi_0|/\MP)^2$ & \eqref{eq:TReff}
\\
$T_p$ & Radiation transition temperature (inflaton $\to$ Q-ball) & \eqref{eq:Tp}
\\
$Y_B$ & Baryon asymmetry & \eqref{eq:YB_vs_YLa},\eqref{eq:YLa_before_EWPT}
\\ \hline
\end{tabular}
\caption{List of input parameters and key derived quantities}
\label{tab:list_of_paras}
\end{table}

\section{Effective number of Q-ball decay channels}
\label{app:Ndof}

In this Appendix, we evaluate the effective number of decay channels contributing to the Q-ball charge emission, $N$ in Eq.~\eqref{eq:Gamma_Q}.

For a decay mode into a one-degrees-of-freedom fermion, the maximum production rate of the fermion is~\cite{Kawasaki:2012gk}
\begin{align}
\frac{dN_{\text{1dof}}}{dt}
\le
\left. 
\frac{dN_{\text{1dof}}}{dt}
\right|_{\text{sat.}}
\simeq
\frac{\omega_Q^3}{12\pi^2} 4\pi R_Q^2
\end{align}
The saturation rate, independent of the coupling, is achieved due to the Pauli exclusion principle as far as the field value inside the Q-ball is sufficiently large~\cite{Kawasaki:2012gk}, which is the case in our scenario. The dominant mode is an annihilation like $\phi \phi \to qq, \ell\ell$, where $q$ and $\ell$ collectively denote quarks and leptons, respectively. 
Therefore, the production rates of the leptons with flavor $\alpha$ ($\alpha=e,\mu,\tau$) are bounded as
\begin{align}
\frac{dN_{L_\alpha}}{dt}
\le
\left. 
\frac{dN_{L_\alpha}}{dt}
\right|_{\text{sat.}}
=
g_\alpha
\left. 
\frac{dN_{\text{1dof}}}{dt}
\right|_{\text{sat.}}.
\end{align}
where $g_\alpha=3$ is the degrees of freedom for the flavor $\alpha$, which accounts for left- and right-handed charged lepton and the neutrino. Q-ball can also decay into quarks, with the rate
\begin{align}
\frac{dN_q}{dt}
\le
\left. 
\frac{dN_q}{dt}
\right|_{\text{sat.}}
=
g_{\text{quarks}}
\left. 
\frac{dN_{\text{1dof}}}{dt}
\right|_{\text{sat.}}.
\end{align}
with $g_{\text{quarks}}$ being the degrees of freedom for quarks, including color, flavor, and helicity. As we will see shortly, its precise value does not matter as far as $g_{\text{quarks}}>g_\alpha$. 

Let us consider the lepton with the flavor $\alpha$. The total Q-ball charge $Q$ is related to the $L_\alpha$ number $N_{L_\alpha}$ as $N_{L_\alpha}=(\kappa_\alpha/4)Q$. (See Eq.~\eqref{eq:nLi_nphi}.) This relation should hold during the Q-ball decay, as far as the $D$-flat condition is satisfied. Therefore, from the above equation, 
\begin{align}
\left|
\frac{dQ}{dt}
\right|
=
\frac{4}{|\kappa_\alpha|}\frac{dN_{L_\alpha}}{dt}
\le
\frac{4}{|\kappa_\alpha|} g_\alpha
\left. 
\frac{dN_{\text{1dof}}}{dt}
\right|_{\text{sat.}}.
\end{align}
Similarly, for quarks, 
\begin{align}
\left|\frac{dQ}{dt}\right|
=
4\frac{dN_q}{dt}
\le
4g_{\text{quarks}}
\left. 
\frac{dN_{\text{1dof}}}{dt}
\right|_{\text{sat.}}.
\end{align}
The Q-ball decay rate is saturated by the slowest mode among the above ones, which is the production of the $\beta$-flavor lepton with $|\kappa_\beta|=1$.
This results in
\begin{align}
\left|\frac{dQ}{dt}\right|
=
4\frac{dN_{L_\beta}}{dt}
=
4g_\beta
\left. 
\frac{dN_{\text{1dof}}}{dt}
\right|_{\text{sat.}}
=
12
\left. 
\frac{dN_{\text{1dof}}}{dt}
\right|_{\text{sat.}},
\end{align}
and hence the effective number of decay channels, $N$, in Eq.~\eqref{eq:Gamma_Q} is given by $N=12$.
The other modes, i.e., decays into lepton flavor $\alpha$ with $|\kappa_\alpha| < 1$ as well as into quarks, are suppressed because the decay into the $\beta$-flavor lepton is rate-limiting.

\section{Contributions from Q-ball evaporation}
\label{app:evap-diff}

In this Appendix, we show that the evaporation and the diffusion of Q-balls give negligible contributions to the baryon asymmetry in most of the parameter regions of our interest.
We also briefly comment on the possible suppression of the Q-ball decay.

At high temperatures,
Q-balls can emit their charges 
through evaporation, with the emission rate given by~\cite{Laine:1998rg}
\begin{align}
    \Gamma_{\rm evap}\equiv \frac{dQ}{dt}\biggl|_{\rm evap}&\simeq  -4\pi\xi^\prime \left(\mu_Q-\mu_{\rm plasma}\right)T^2R_Q^2
    \label{eq:Gamma_evap_1}
\end{align}
where $\mu_Q\simeq \omega_Q$ and $\mu_{\rm plasma}$ are the chemical potentials for Q-balls and plasma, respectively. 
In the following, for simplicity, we discuss the temperature regions of $T>T_*$ and $T<T_*$ separately with $T_*\simeq 500~{\rm GeV}$. We assume $T_* > T_p$, which is satisfied for $T_D\lesssim 100~{\rm GeV}(T_*/500~{\rm GeV})^{5/4}(T_{R,{\rm eff}}/10^5~{\rm GeV})^{-1/4}$.

\begin{itemize}

\item $T>T_*$: 
Since the evaporation rate \eqref{eq:Gamma_evap_1} is suppressed as $\mu_{\rm plasma}$ approaches $\mu_Q$, the plasma chemical potential is bounded above by $\mu_{\rm plasma}\le\mu_Q$. This implies an upper bound on the lepton flavor chemical potential: $\mu_\alpha\simeq \mu_{\rm plasma}\lesssim \mu_Q\simeq \omega_Q$. 
Assuming $\mu_\alpha < c\ \omega_Q$ with $c\sim O(1)$, the saturated lepton flavor asymmetry evaluated at a temperature $T=T_*$ is given by
\begin{align}
\left|
Y_{L_\alpha}(T_*)
\right|
<
\left|
Y_{L_\alpha}^{\rm sat}(T_*)
\right|
&=
\frac{1/6}{2\pi^2 g_\ast/45} 
\frac{c\ \omega_Q}{T_*},
\label{eq:YL_sat}
\end{align}
where we have used Eq.~\eqref{eq:YL_vs_xi} and omitted the $\xi^3$ term for simplicity.
The corresponding baryon asymmetry is then given by
\begin{align}
\left|
Y_B^{\rm sat}\right|_{T_D}
&\lesssim
0.030 \sum_{\alpha}\ h_\alpha^2
\left|
Y_{L_\alpha}^{\rm sat}(T_*)
\right|
\frac{3T_D}{T_{R,{\rm eff}}}
\end{align}
where we have used Eq.~\eqref{eq:YB_vs_YLa}, and the last factor $3T_D/T_{R,{\rm eff}}$ accounts for the entropy dilution factor.
Using Eqs.~\eqref{eq:omega_Q}, we find that the contribution to the baryon asymmetry from  the evaporation at $T>T_*$ is bounded from above as
\begin{align}
\left|
Y_B^{{\rm evap}, T>T_*}
\right|
<
\left|
Y_B^{\rm sat}\right|_{T_D}
\simeq
4\times 10^{-13}
\left(\frac{T_*}{500~{\rm GeV}}\right)^{-1}
\left(\frac{m_{3/2}}{1~{\rm GeV}}\right)
\left(\frac{T_D}{4~{\rm GeV}}\right)
\left(\frac{T_{R,{\rm eff}}}{10^5~{\rm GeV}}\right)^{-1}
c\
\widehat{\beta}^{-1/4}
\widehat{\zeta}
\label{eq:YB_sat}
\end{align}
where we have used $g_*(T_*)\simeq 100$.
Therefore, we can see that Q-ball evaporation at $T>T_*$ gives a negligible contribution to the baryon asymmetry compared with the observed one, $Y_B^{\rm obs}\simeq 10^{-10}$, for most of the allowed parameter regions in Fig.~\ref{fig:parameterspace_L}.

At even higher temperatures, the diffusion may become relevant~\cite{Banerjee:2000mb}. Its rate is smaller than Eq.~\eqref{eq:Gamma_evap_1}, and hence its contribution to the final baryon asymmetry is also subdominant.

\item $T\le T_*$: At low temperatures, the saturation $\mu_Q\simeq \mu_{\rm plasma}$ may not happen. Let us estimate the maximum baryon asymmetry from the Q-ball evaporation at $T\le T_*$, assuming $\mu_Q\gg \mu_{\rm plasma}$. In this case, the evaporation rate in Eq.~\eqref{eq:Gamma_evap_1} is given by~\cite{Banerjee:2000mb}\footnote{We have adopted the formula in Ref.~\cite{Banerjee:2000mb}, $\Gamma_{\rm evap}\simeq -\kappa' 4\pi R_Q^2 n_B^{\rm eq}$ with $\kappa'\simeq 1$ for $T\gg m_\phi$ and $n_B^{\rm eq}=(p g_B /6) (M_Q/Q) T^2$ with $p=3/4$, replaced $g_B$ with $N$ introduced in Eq.~\eqref{eq:Gamma_Q}, and used Eq.~\eqref{eq:Q-ball-properties}. 
}
\begin{align}
    \Gamma_{\rm evap}
    &\simeq
    -4\pi \xi R_Q^2 (N/6) \omega_Q T^2
    \quad 
    (\mu_Q\gg \mu_{\rm plasma}
)    
    \label{eq:Gamma_evap}
\end{align}
where the coefficient $\xi$ is suppressed as $\xi\simeq \widehat{\xi}(T/m_\phi)^2$ for $T \le  T_* <  m_\phi$~\cite{Laine:1998rg,Banerjee:2000mb},\footnote{$T_p<m_\phi$ is always satisfied in the parameter ranges of our interest.} and we assume $\widehat{\xi}\simeq 1$, for simplicity.\footnote{For $T\ll m_\phi$, $\widehat{\xi}$ can be suppressed as $\widehat{\xi}\sim \alpha_W^2$ with $\alpha_W$ being the electroweak coupling~\cite{Kawasaki:2002hq}. Here, we assume $\widehat{\xi}\simeq 1$ for simplicity, which gives a conservative bound on the evaporation's contribution.}
Using the Hubble-temperature relation in Eqs.~\eqref{eq:T_from_inf}, \eqref{eq:T_from_Q} and $H=2/(3t)$ in the Q-ball dominated Universe, and by integrating $dQ/dt=\Gamma_{\rm evap}$, we obtain
\begin{align}
\Delta Q_{\rm evap}(T<T_*) &\lesssim
\Delta Q_{\rm evap}^{\rm max}(T<T_*) 
\simeq
\frac{4\pi^3 N \widehat{\xi}}{15}
\left(\frac{\pi^2 g_*(T_*)}{270}\right)^{-1/2}
\frac{\MP }{\omega_Q m_\phi^2 T_{R,{\rm eff}}^{1/2}}
T_*^{5/2}.
\end{align}
where the inequality is saturated if $\mu_Q\gg \mu_{\rm plasma}$ for $T\lesssim T_*$.
This results in a lepton flavor asymmetry
\begin{align}
\left|
Y_{L_\alpha}^{{\rm evap},T\le T_*}
\right|_{T_D}
&\lesssim 
\frac{\Delta Q_{\rm evap}^{\rm max}(T\le T_*)}
{Q_i} 
\left|Y_{L_\alpha}\right|,
\end{align}
where $Q_i$ and $Y_{L_\alpha}$ are given by Eq.~\eqref{eq:Qinit} and Eqs.~\eqref{eq:YL_withQ_1},\eqref{eq:YL_withQ_2}, respectively. The corresponding baryon asymmetry is given by
\begin{align}
\left|Y_B^{{\rm evap}, T<T_*}\right|_{T_D}
&\lesssim 
\frac{\Delta Q_{\rm evap}^{\rm max}(T<T_*)}{Q_i} 
\times
0.030
\sum_{\alpha}
h_\alpha^2
\left|Y_{L_\alpha}\right|
\notag\\
&\lesssim
7\times 10^{-13}
\left(\frac{m_{3/2}}{1~{\rm GeV}}\right)^{9/2}
\left(\frac{M_F}{10^6~{\rm GeV}}\right)^{-6}
\notag\\
&\times
\kappa_\alpha
\left(\frac{\delta_{\rm CP}\lambda}{0.1}\right)
\left(\frac{m_\phi}{10^4~{\rm GeV}}\right)^{-2}
\left(\frac{|\phi_0|}{\MP}\right)
\left(\frac{T_*}{500~{\rm GeV}}\right)^{5/2}
\left(\frac{T_R}{10^5~{\rm GeV}}\right)^{-1/2}
\widehat{\beta}^{-11/8}
\widehat{\zeta}^{-1/2}
\widehat{N}^{3/2}
\widehat{\xi},
\label{eq:YB_evap_max}
\end{align}
for $\kappa_\tau\ne 0$, and further multiplied by $(h_\mu/h_\tau)^2\simeq 4\times 10^{-3}$ for $\kappa_\tau = 0$.
Therefore, again, we conclude that the contribution from the Q-ball evaporation at $T<T_*$ is negligible in most of the allowed region in Fig.~\ref{fig:parameterspace_L}.
\end{itemize}
Therefore, we conclude that the
evaporation and the diffusion give negligible contributions to the baryon asymmetry in the allowed regions in Fig.~\ref{fig:parameterspace_L}.

Finally, let us comment on the possible suppression of the Q-ball decay by the saturation discussed above. 
Since the Q-ball decay is a non-thermal process and the above discussion may not apply, we do not consider the whole parameter range but just briefly discuss the benchmark point in Sec.~\ref{sec:implication}. At the point, $m_{3/2}\simeq 7$ GeV, and $\omega_Q\simeq 1100~{\rm GeV}$. At $T=T_{\rm sph}\simeq 130~{\rm GeV}$, under the assumption of the saturation, it leads to $\xi_\alpha^{\rm sat}\simeq \omega_Q/T_{\rm sph}\simeq 8.2$ and $Y^{\rm sat}_{L_\alpha}(T_{\rm sph})\simeq 0.24$
(cf. Eq.~\eqref{eq:YL_vs_xi}).
Using $T_D\simeq 17~{\rm GeV}$ and $T_{R,{\rm eff}}\simeq 10^5~{\rm GeV}$, the entropy dilution factor is given by $3T_D/T_{R,{\rm eff}}\simeq 5\times 10^{-4}$. Combined with the sphaleron conversion factor in Eq.~\eqref{eq:YB_vs_YLa}, we find that $Y_B^{\rm sat}\simeq 4\times 10^{-10}$, which is larger than the actual baryon asymmetry obtained in Sec.~\ref{sec:implication}. Therefore, the effect of the suppression from the saturation can be safely neglected.

\section{Gravitinos from Q-ball decay}
\label{app:Gravitino_from_Q}

In this Appendix, we estimate the gravitino abundance from Q-ball decay.
The branching fraction of the Q-ball decay into the gravitino for $\omega_Q<M_{\tilde{g}}$ is~\cite{Kasuya:2012mh}, 
\begin{align}
B_{3/2}\simeq \frac{f_{3/2}^2}{f_q^2},
\end{align}
where
\begin{align}
f_{3/2} \simeq \frac{\omega_Q^2}{\sqrt{3} m_{3/2} \MP},
\quad
f_q \simeq \frac{M_{\tilde{g}}}{\phi_Q},
\quad
\phi_Q &\simeq \frac{1}{\sqrt{2}}\zeta M_F Q^{1/4}.
\end{align}
Using the Eqs.~\eqref{eq:Qinit} and \eqref{eq:omega_Q}, we obtain
\begin{align}
B_{3/2}\simeq \frac{2 \pi^4 \zeta^6
M_F^4 }{3 \beta^{1/2} M_{\tilde{g}}^2 M_P^2 },
\end{align}
which leads to
\begin{align}
\frac{\rho_{3/2}}{s}
&\simeq \frac{n_\phi}{s} B_{3/2} m_{3/2}
\\
&\simeq 4 \left|Y_{L_\alpha}\right|_{\kappa_\alpha=1}  B_{3/2} m_{3/2},
\end{align}
and
\begin{align}
\Omega_{3/2}h^2
&\simeq \frac{\rho_{3/2}/s}{\rho_{\rm crit} h^{-2} /s}
\\
&\simeq 1.1\times 10^{-6}
\left(\frac{m_{3/2}}{1~{\rm GeV}}\right)^{5/2}
\left(\frac{M_F}{10^6~{\rm GeV}}\right)^2
\left(\frac{M_{\tilde{g}}}{10^4~{\rm GeV}}\right)^{-2}
\left(\frac{\delta_{\rm CP}\lambda}{0.1}\right)
\left(\frac{|\phi_0|}{\MP}\right)^2
\widehat{\beta}^{-9/8}
\widehat{\zeta}^{13/2}
\widehat{N}^{1/2},
\end{align}
where we have used Eq.~\eqref{eq:YL_withQ_2} and $\rho_{\rm crit} h^{-2}/s=3.64\times 10^{-9}~{\rm GeV}$.

\bibliography{bib.bib}

\end{document}